\documentclass[letterpaper,titlepage,11pt]{article}

\pdfoutput=1

\usepackage{amssymb,amsmath,amsfonts, graphics}
\usepackage{epsfig}
\usepackage{ccaption}
\usepackage{graphicx}

\usepackage[
      colorlinks=true,
      linkcolor=blue,
      urlcolor=blue,
      filecolor=black,
      citecolor=red,
      pdfstartview=FitV,
      pdftitle={},
        pdfauthor={Roberto Emparan, Ryotaku Suzuki, Kentaro Tanabe},
        pdfsubject={},
        pdfkeywords={},
        pdfpagemode=None,
        bookmarksopen=true
      ]{hyperref}

\def\clock{{\count0=\time
           \divide\count0 60
           \ifnum\count0<10 0\fi\the\count0
           \multiply\count0 -60 \advance\count0 \time
           :\ifnum\count0<10 0\fi \the\count0
         }}
\newcommand{\timestamp}{{\small\vbox{\hbox{\tt\jobname.tex}
\hbox{\the\day/\the\month/\the\year, \clock}}}}

\newcommand{\ie}{{\it i.e.,\,}}

\newcommand{\lp}{\left(}
\newcommand{\rp}{\right)}

\newcommand{\beq}{\begin{equation}}
\newcommand{\eeq}{\end{equation}}
\newcommand{\bea}{\begin{eqnarray}}
\newcommand{\eea}{\end{eqnarray}}
\newcommand{\beqa}{\begin{eqnarray}}
\newcommand{\eeqa}{\end{eqnarray}}

\newcommand{\sR}{\mathsf{R}}

\newcommand{\Or}{\mathcal{O}}

\begin{document}

\begin{titlepage}
\rightline{KEK-TH-1838, OCU-PHYS-429, AP-GR-126} 
\leftline{}
\vskip 2cm
\centerline{\LARGE \bf Evolution and endpoint of the black string instability:}
\medskip
\centerline{\LARGE \bf Large $D$ solution} 
\vskip 1.6cm
\centerline{\bf Roberto Emparan$^{a,b}$, Ryotaku Suzuki$^{c}$, Kentaro Tanabe$^{d}$}
\vskip 0.5cm
\centerline{\sl $^{a}$Instituci\'o Catalana de Recerca i Estudis
Avan\c cats (ICREA)}
\centerline{\sl Passeig Llu\'{\i}s Companys 23, E-08010 Barcelona, Spain}
\smallskip
\centerline{\sl $^{b}$Departament de F{\'\i}sica Fonamental, Institut de
Ci\`encies del Cosmos,}
\centerline{\sl  Universitat de
Barcelona, Mart\'{\i} i Franqu\`es 1, E-08028 Barcelona, Spain}
\smallskip
\centerline{\sl $^{c}$Department of Physics, Osaka City University, Osaka 558-8585, Japan}
\smallskip
\centerline{\sl $^{d}$Theory Center, Institute of Particles and Nuclear Studies, KEK,}
\centerline{\sl  Tsukuba, Ibaraki, 305-0801, Japan}
\smallskip
\vskip 0.5cm
\centerline{\small\tt emparan@ub.edu,\, ryotaku@sci.osaka-cu.ac.jp,\, ktanabe@post.kek.jp}

\vskip 1.6cm
\centerline{\bf Abstract} \vskip 0.2cm \noindent
We derive a simple set of non-linear, $(1+1)$-dimensional partial differential equations that describe the dynamical evolution of black strings and branes to leading order in the expansion in the inverse of the number of dimensions $D$. These equations are easily solved numerically. Their solution shows that thin enough black strings are unstable to developing inhomogeneities along their length, and at late times they asymptote to stable non-uniform black strings. This proves an earlier conjecture about the endpoint of the instability of black strings in a large enough number of dimensions. If the initial black string is very thin, the final configuration is highly non-uniform and resembles a periodic array of localized black holes joined by short necks. We also present the equations that describe the non-linear dynamics of Anti-deSitter black branes at large $D$.

\end{titlepage}
\pagestyle{empty}
\small
\normalsize
\newpage
\pagestyle{plain}
\setcounter{page}{1}




\paragraph{I.} The instability of black strings and black branes discovered in \cite{Gregory:1993vy} is a phenomenon with wide implications for the physics of higher-dimensional black holes and their applications to string theory and gauge/gravity duality \cite{GLreview}. Black strings behave similarly to tubular soap films and are prone, when thin enough, to rippling, \ie developing non-uniformities along their direction. This instability is well established in perturbation theory, but its growth beyond the linearized approximation and its endpoint at asymptotically late times are notoriously hard problems. The numerical evolution of a perturbed five-dimensional black string in \cite{Lehner:2010pn} is a landmark result: it gives strong evidence that the classical evolution does not stop at any stable configuration but proceeds in a self-similar cascade to smaller scales. 

Important as the result of \cite{Lehner:2010pn} is, it still leaves open many questions about the fate of unstable black strings and black branes. For instance, it would be very convenient to have a better, possibly analytic, understanding of the late-time evolution. Moreover, \cite{Lehner:2010pn} presents a single calculation of a single system. Do all unstable black strings behave in the same manner? This is indeed unlikely: there has long been evidence that, by modifying the parameters of the system, the endpoint may be different. In the simplest instance, the parameter is the number of spacetime dimensions where the black string lives. Ref.~\cite{Sorkin:2004qq} found a critical dimension, $D=D_*\simeq 13.5$, above which weakly non-uniform static  black strings (NUBS) have larger horizon area, for fixed string length and mass, than the uniform solutions. It is then possible that the classical evolution of the system in $D>D_*$ ends at a stable NUBS, as proposed (independently of $D$) in \cite{Horowitz:2001cz}. However, given the almost prohibitive cost of the numerical simulations ($100\, 000$ CPU hours for \cite{Lehner:2010pn}), the investigation of this possibility and the systematic study of the evolution of related systems has not been undertaken to this date. This is strong motivation to search for simpler methods capable of capturing at least the main qualitative features of the phenomenon.

In this article we present an approach, based on an expansion in $1/D$, that allows to address some of these issues. In particular, we show that for large enough $D$ the endpoint of the instability is generically a stable NUBS. The simplification of the problem is dramatic: accurate numerical evolutions can be obtained in seconds (or less) in a conventional computer running a one-line \texttt{NDSolve} of \textsl{Mathematica}. This encourages further investigation of this and similar problems by these means. 

\paragraph{II.} The large-$D$ approach to black hole physics, initiated in \cite{Asnin:2007rw,Emparan:2013moa,Emparan:2013xia}, has been recently developed to deal with fully non-linear deformations of horizons \cite{Emparan:2015hwa,Bhattacharyya:2015dva,Suzuki:2015iha,Suzuki:2015axa}. While the formalism of \cite{Bhattacharyya:2015dva} incorporates time evolution, it is unclear to us whether it allows, as it is, to study the system at hand, which involves horizon length scales $\sim 1/\sqrt{D}$. Therefore, we solve the problem \textit{ab initio} in a formulation adapted to the dynamics of black branes.

Using ingoing Eddington-Finkelstein coordinates, the metric of a uniform black $p$-brane in $D=n+p+3$ dimensions, boosted along its worldvolume with velocity vector $u_\mu$, $\eta^{\mu\nu}u_\mu u_\nu=-1$, $\mu,\nu=0,\dots,p$, is
\beq\label{ubsEF}
ds^2=\lp \eta_{\mu\nu}+\lp \frac{r_0}{r}\rp^n u_\mu u_\nu\rp d\sigma^\mu d\sigma^\nu -2u_\mu d\sigma^\mu dr +r^2d\Omega_{n+1}\,.
\eeq 
The worldvolume directions are 
\beq
\sigma^\mu=(t, \sigma^a)\,,\quad a=1,\dots,p\,.
\eeq
We take $p$ to be finite, so $D\to\infty$ is $n\to\infty$, and we will study deformations of the black brane that depend on $\sigma^\mu$.

Since the wavelength of unstable fluctuations is known to be $\sim 1/\sqrt{n}$ \cite{Asnin:2007rw,Emparan:2013moa}, we rescale $\sigma^a\to \sigma^a/\sqrt{n}$. Furthermore, we consider small velocities,
$
u_\mu=\lp -1+\Or\lp n^{-1}\rp, v_a/\sqrt{n}\rp,
$
so, denoting $m=r_0^n$, the metric \eqref{ubsEF} becomes
\beqa\label{UBS2}
ds^2&=&-\lp 1-\frac{m}{r^n}\rp dt^2  +2 \lp dt -\frac{v_a}{n}d\sigma^a\rp dr-\frac{2m v_a}{n\, r^n}d\sigma^a dt\notag\\
&&+\frac{1}{n}\lp \delta_{ab}+\frac{m v_a v_b}{n\, r^n}\rp d\sigma^a d\sigma^b+r^2d\Omega_{n+1}\,.
\eeqa

We seek solutions that can be regarded as having $m$ and $v_a$ not constant but varying with the coordinates $\sigma^\mu$. In order to find them, we take a metric ansatz of the form
\beqa
ds^2=-Adt^2-2(u_t dt+u_a d\sigma^a)dr - 2C_a d\sigma^a dt +G_{ab}d\sigma^a d\sigma^b +r^2d\Omega_{n+1}\,,\eeqa
with
\beq
A=\sum_{k\geq 0}\frac{A^{(k)}(\sigma^\mu,r)}{n^k}\,,\quad
u_t=\sum_{k\geq 0}\frac{u_t^{(k)}(\sigma^\mu,r)}{n^k}\,,\quad
u_a=\sum_{k\geq 0}\frac{u_a^{(k)}(\sigma^\mu,r)}{n^{k+1}}\,,\quad
\eeq
\beq
C_a=\sum_{k\geq 0}\frac{C_a^{(k)}(\sigma^\mu,r)}{n^{k+1}}\,,\quad
G_{ab}=\frac{1}{n}\lp \delta_{ab}+\sum_{k\geq 0}\frac{G_{ab}^{(k)}(\sigma^\mu,r)}{n^{k+1}}\rp\,.
\eeq
The different scalings with $n$ conform to \eqref{UBS2}. We have chosen a Bondi-type gauge where $g_{rr}=0$ and $r$ is the area-radius of $S^{n+1}$ to all orders in $1/n$. This leaves a gauge freedom in the choice of $u_a$, which is a shift vector on surfaces at constant $r$ and is only restricted by boundary conditions. We have partially gauge-fixed it to be independent of $r$. 

We introduce the radial coordinate
\beq
\sR =\lp \frac{r}{r_0}\rp^n\,,
\eeq
such that when $D\to\infty$ keeping $\sR$ finite, we focus on the near-horizon region.
The boundary conditions at large $\sR$ (for near-horizon decoupled geometries \cite{Emparan:2014aba}) are
\beq
A=1+\Or(\sR^{-1})\,,\qquad C_a=\Or(\sR^{-1})\,,\qquad G_{ab}=\frac{1}{n}\lp \delta_{ab}+\Or\lp n^{-1},\sR^{-1}\rp\rp\,.
\eeq
We also require regularity at the horizon, which is where $g_{\mu\nu}u^\mu u^\nu=0$. Henceforth we fix $r_0=1$.

It is now straightforward to solve the Einstein equations perturbatively in $1/n$. To leading order, the solution is 
\beq
A^{(0)}=1-\frac{m(\sigma^\mu)}{\sR}\,,\qquad C_a^{(0)}=\frac{p_a(\sigma^\mu)}{\sR}\,,\qquad G_{ab}^{(0)}=\frac{p_a(\sigma^\mu) p_b(\sigma^\mu)}{m(\sigma^\mu)\sR}\,,
\eeq
\beq
u^{(0)}_t=-1\,,\qquad u^{(0)}_a=\textrm{constant}\,.
\eeq
We have gauge-fixed the $\sigma^\mu$-dependence of $u_a$ in a manner consistent with boundary conditions. 
Solving similarly at the next order, the equations $R_{tt}=0$ and $R_{ta}=0$ are $r$-independent constraints that require, respectively,
\beq\label{eqm}
\partial_t m-\partial_b\partial^b m=-\partial_b p^b\,,
\eeq
and
\beq\label{eqp}
\partial_t p_a-\partial_b\partial^b p_a=\partial_a m-\partial_b \lp \frac{p_a p^b}{m}\rp\,
\eeq
($a,b$ indices are raised with the flat metric $\delta^{ab}$). These equations are one of our main results. They are the effective equations for the collective variables $m(\sigma^\mu)$, $p_a(\sigma^\mu)$ that describe non-linear fluctuations of the black brane.
These variables give the energy and momentum densities of the black brane on sections at constant $\sigma^a$ at a given time. 
The horizon of the solution is at $\sR=m(\sigma^\mu)$, so the function $m(\sigma^\mu)$ is also interpreted as the area-density, \ie the area of the $S^{n+1}$ (up to a factor of the unit-sphere area) at the horizon at a fixed value of $t$ and $\sigma^a$, to leading order in $1/n$. 
For solutions that are spatially periodic in $\sigma^a$, the quantities
\beq\label{MP}
M=\int d\sigma^1\dots d\sigma^p\, m(t,\sigma^b)\,,\qquad P_a=\int d\sigma^1\dots d\sigma^p\, p_a(t,\sigma^b)\,,
\eeq
are conserved in time. Up to normalization factors, they are the total mass and momentum of the black brane. Up to similar factors and to leading order, $M$ is also the total horizon area, which therefore does not vary as the system evolves. If we define the surface gravity as the non-affinity of the horizon generator $\partial_t$, it also remains constant.

The equations are invariant under Galilean boosts along $\sigma^a$ with constant velocity $v_a$,
\beq
\sigma^a\to \sigma^a- v_a t\,,\qquad p_a\to p_a +m v_a\,,
\eeq
which allows to fix the rest frame of the black brane, in which $P_a=0$, and set $u^{(0)}_a=0$.

This effective formulation of black brane dynamics passes two important checks: (i) For small, linearized perturbations around the uniform black brane, with momentum $k$ aligned with $\sigma^1\equiv z$,
\beq\label{pert}
m(t,z)=1+\delta m\, e^{-i\omega t+ik z}\,,\qquad p_a(t,z)=\delta p_a\, e^{-i\omega t+ik z}\,,
\eeq
the solution frequencies are
\beq\label{linmodes}
\omega_\pm =i(\pm k-k^2)\,,\qquad \omega =-ik^2\,,
\eeq
which reproduce the frequencies of the sound and shear modes of the black brane to leading order at large $D$ \cite{Emparan:2013moa}. In particular, the frequency $\omega_+$ for $0<k<1$ corresponds to the Gregory-Laflamme unstable mode. (ii) If we consider static, shear-free deformations, $m=m(z)$, $p_1=p(z)$, $p_{a\neq 1}=0$, the resulting equation (with $p=m'$ in the rest frame) 
\beq\label{stateq}
m''+m-\frac{(m')^2}{m}=\mathrm{constant},
\eeq
is equivalent to the equation for static black strings derived in \cite{Emparan:2015hwa,Suzuki:2015axa} (using different gauges), with the variables being related by $m(z)=\exp(2\mathcal{P}(z))$. 

Observe that the sound deformations $m$, $p_1$ along a direction $\sigma^1$, are not affected by the shear $p_{a\neq 1}$. In the following we set $p_{a\neq 1}=0$ and consider black strings with $p_1(t,\sigma^1=z)\equiv p(t,z)$. The direction $z$ is compactified, $z\in [-L/2,L/2]$. We parametrize the periodicity $L$ in terms of a wavenumber $k_L$ as
\beq
k_L=\frac{2\pi}{L}\,.
\eeq
Since we are fixing the string thickness $r_0=1$, the uniform strings are characterized by the value of $k_L$. Smaller values of $k_L$ correspond to thinner black strings.

\paragraph{III.} We solve numerically the black string equations
\beq\label{eqm2}
\partial_t m(t,z)-\partial_z^2 m(t,z)=-\partial_z p(t,z)\,,
\eeq
and
\beq\label{eqp2}
\partial_t p(t,z)-\partial_z^2 p(t,z)=\partial_z \lp m(t,z)-\frac{p(t,z)^2}{m(t,z)}\rp\,.
\eeq
The \textsl{Mathematica} function \texttt{NDSolve} handles them without difficulty. We fix a value of $k_L$ and introduce a small perturbation of the static uniform black string, $m(0,z)=1+\delta m_0(z)$, $p(0,z)=\delta p_0(z)$, such that the momentum $P$ vanishes. We find that for $k_L>1$, the perturbation quickly dissipates and the black string becomes uniform, in agreement with the absence in \eqref{linmodes} of unstable linear modes with wavenumber $k>1$.

For thinner black strings, with $k_L<1$, after brief initial transients the deformation grows (at approximately the exponential rate of the linearized solution). Eventually, the system settles down at a stable configuration that approximates very well a solution of the static equation \eqref{stateq}. $M$ and $P$ in \eqref{MP} remain constant to very good accuracy throughout the evolution. 

\begin{figure}[th]
\begin{center}
\includegraphics[width=15cm]{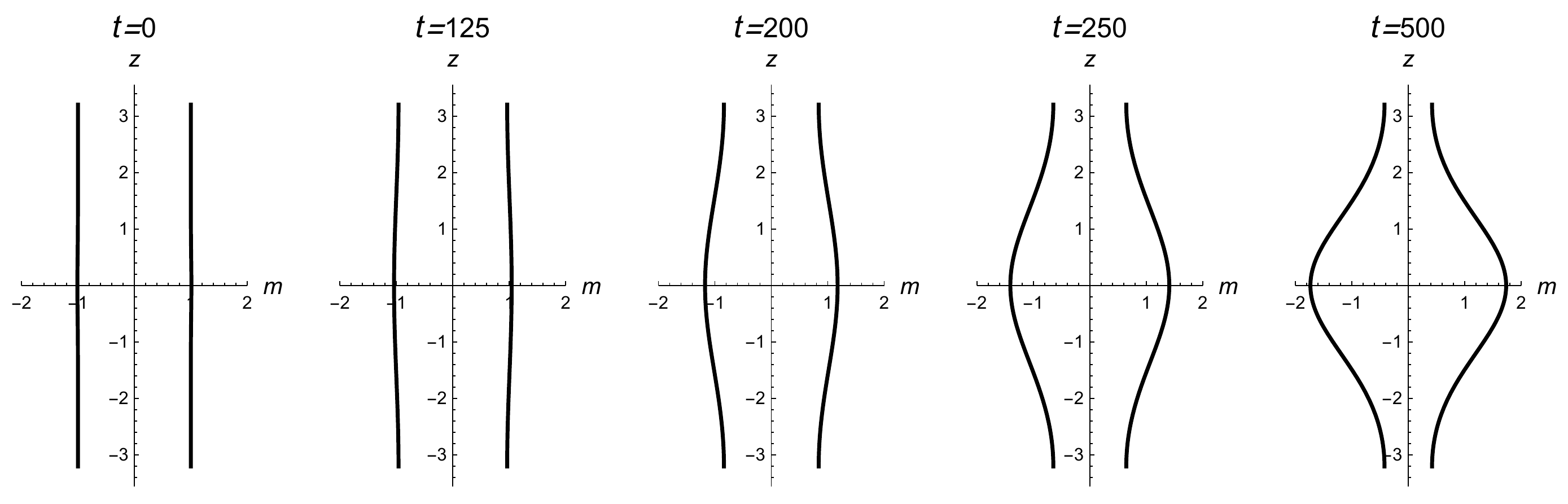}
\caption{\small Dynamical evolution of a perturbed string with $k_L=0.98$. The horizontal axis is the $S^{n+1}$-area function $m(t,z)$. For this simulation, the final state was reached before $t=500$. The time it takes depends on the size and shape of the initial perturbation, but the final configuration is independent of them.}  
\label{fig:k098}
\end{center}
\end{figure}

\begin{figure}[th]
\begin{center}
\includegraphics[width=16cm]{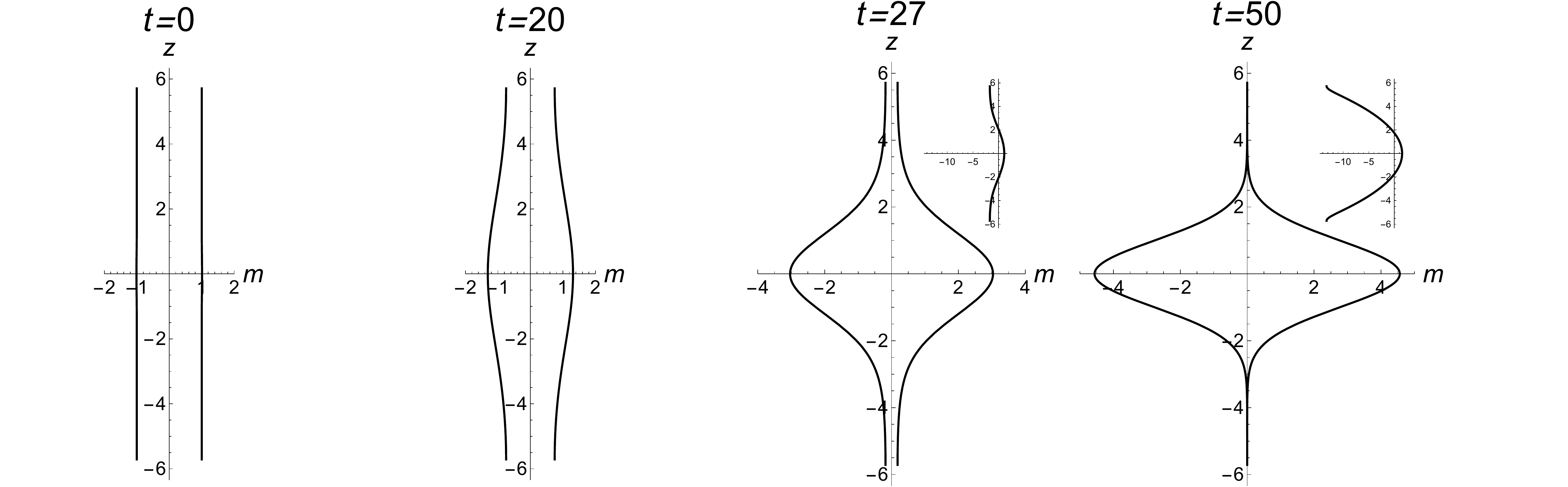}
\caption{\small Dynamical evolution of a perturbed string with $k_L=0.55$. For the last two plots we display  $\ln m(t,z)$ in insets, which show that the blobs extend to almost fill the compact direction. For this simulation, the final state was reached around $t=40$.}  
\label{fig:k055}
\end{center}
\end{figure}

In figures \ref{fig:k098} and \ref{fig:k055} we show two sample simulations. We plot $m(t,z)$, which gives the area of the $S^{n+1}$ at a given $z$. This is different than the area-radius of these spheres, which is
\beq
\mathcal{R}(t,z)=m(z)^\frac{1}{n+1}\simeq 1+\frac{\ln m(t,z)}{n}\,.
\eeq
For large deformations, $\ln m(t,z)$ can reveal structure that in $m(t,z)$ is exponentially suppressed (insets in fig.~\ref{fig:k055}).

Fig.~\ref{fig:k098} is the evolution of a not-too-thin black string with $k_L=0.98$. Since the perturbative unstable rate is small, the evolution is slow. The final profile is approximately sinusoidal, as expected for a small deformation.

Fig.~\ref{fig:k055} shows the evolution for a thinner black string with $k_L=0.55$, which evolves much faster and develops a large blob in its final state. As shown in \cite{Emparan:2015hwa,Suzuki:2015axa}, the profile of large static blobs is very approximately gaussian,
\beq\label{bulge}
m(z)\simeq \frac{L}{\sqrt{2\pi}} e^{-z^2/2}
\eeq
(the amplitude is fixed to have the same $M$ as a uniform black string of length $L$). The radius $\mathcal{R}(z)$ for \eqref{bulge} decays only like $-z^2/2$ away from the center, as seen in the insets in fig.~\ref{fig:k055}, and the blob fills almost all the compact direction. Indeed, \eqref{bulge} gives an excellent approximation to the profile of $\ln m(z)$ for the final solution at all $z$ (better than $1\%$ for $k_L\lesssim 0.6$) except in a very short `neck' of length $(\Delta z)_\mathrm{neck}\simeq (4/L)\ln L$ near $|z|=L/2$ \cite{Emparan:2015hwa}. As shown in \cite{Suzuki:2015axa}, these blobs approximate very well the shape of a spherical black hole within a region of width $\sim 1/\sqrt{n}$ around its equator, as well as its total area. Bear in mind, however, that while the height of the blobs is $\mathcal{R}=\Or(1)$, their proper length along $z$ is $L/\sqrt{n}$, \ie\ they are much thinner than a sphere.

When $1/2<k_L<1$ the NUBS at the endpoint of the evolution is unique for each value of $k_L$, independently of the shape and amplitude of the initial perturbation. When $0<k_L<1/2$ there are (at least) two different unstable modes that can be excited, with wavenumbers $k_L$ and $2k_L$ (or a higher multiple), and two possible final static NUBS. Since their mass and area are the same to leading order in $1/n$, the evolution depends on the relative growth rates of the unstable modes and on the shape of the initial perturbation. 
When the solution develops two or more blobs, it may be difficult to determine if these will remain in the asymptotic final state, or are instead only part of a very long-lived transient phase: since $m$ is exponentially small inbetween the blobs, the interaction among them may be lost in numerical error. 
These situations, however, may fall outside the range of applicability of this formulation. 

The large-$D$ approximation is valid when $|\ln m|,\, |\partial_{t,z}\ln m|, |\partial_{t,z}\ln p|\ll n$. In the solutions we consider, the most stringent condition is  $|\ln m| \ll n$, which at the neck becomes,  $L\ll 2\sqrt{2n}$, \ie\ $n k_L^2\gg \pi^2/2$.
Thus our results are quantitatively accurate in a given dimension only for sufficiently large $k_L$, or conversely, for a given $k_L$ only in sufficiently large $n$. However, it may be reasonable to expect that the model still captures correctly the qualitative physics for $n$ above the critical dimension $n_*=D_*-4\simeq 9.5$. Observe that near the limit of validity, at $L\sim \sqrt{n}$, the blob has proper size $\Or(1)$ in all directions \cite{Emparan:2015hwa}. This suggests that this final state can be regarded as an array of roughly spherical black holes joined by thin, short necks. Additionally, when the neck becomes thin in Planck units, the black string may break up into separate black holes due to quantum gravity effects.

Within this range of validity, we conclude that our results are very strong evidence that the endpoint of the black string instability at large enough $D$ is generically a stable NUBS. 

\paragraph{IV.} There are several possible extensions of our study. With little extra effort we can obtain the equations that describe the leading large-$D$ non-linear dynamics of Anti-deSitter black branes. Either by directly solving the equations as in the previous case, or by applying the AdS/Ricci-flat correspondence of \cite{Caldarelli:2012hy}, we find
\beq\label{eqmads}
\partial_t m-\partial_b\partial^b m=-\partial_b p^b\,,
\eeq
and
\beq\label{eqpads}
\partial_t p_a-\partial_b\partial^b p_a=-\partial_a m -\partial_b\lp \frac{p_a p^b}{m}\rp\,.
\eeq
It is easy to prove that these equations do not have any non-uniform static solutions. Small dynamical perturbations of the uniform configuration, \eqref{pert}, give as solutions the sound and shear quasinormal frequencies of the AdS black brane computed in \cite{Emparan:2015rva}, namely,
\beq\label{linmodesads}
\omega_\pm =\pm k-ik^2\,,\qquad \omega =-ik^2\,.
\eeq
Other quasinormal modes of the black brane have frequencies $\Or(D)$ and do not appear in the near-horizon decoupling limit. Allowing variation in a number $p\sim D$ of horizon directions may involve qualitative changes.

The inclusion of $1/D$ corrections to the equations \eqref{eqm}, \eqref{eqp} is potentially very interesting. It has been shown in \cite{Suzuki:2015axa} that $1/D$ corrections allow to accurately identify the critical dimension $D_*$ below which static NUBS of a given area have higher mass than uniform black strings, so they cannot be the endpoints of the instability \cite{Gubser:2001ac,Wiseman:2002zc}. This suggests that the $1/D$ expansion may be able to reproduce, when $D<D_*$, an evolution  qualitatively similar to the one observed in \cite{Lehner:2010pn}. We hope to report on this in the future.

\section*{Acknowledgements}

RE is supported by FPA2013-46570-C2-2-P, AGAUR 2009-SGR-168 and CPAN CSD2007-00042 Consolider-Ingenio 2010. KT was supported by JSPS Grant-in-Aid for Scientific Research No.26-3387.


\end{document}